\documentclass[aps,pra,amssymb,superscriptaddress,11pt,tightenlines,onecolumn]{revtex4-2}

\usepackage[english]{babel}
\usepackage{graphicx}
\usepackage{dcolumn}
\usepackage{bm}
\usepackage{color}
\usepackage{subfigure}
\usepackage[ansinew]{inputenc}
\usepackage{amsfonts}
\usepackage{amsmath}
\usepackage{amssymb}

\newcommand{\uniba}{Dipartimento Interateneo di Fisica, Universit\'a degli Studi di Bari, I-70126 Bari, Italy}
\newcommand{\infn}{INFN, Sezione di Bari, I-70125 Bari, Italy}
\newcommand{\upo}{Department of Optics, Palack\'y University, 17. listopadu 12, 77146 Olomouc, Czech Republic}
\newcommand{\epfl}{\'Ecole polytechnique f\'ed\'erale de Lausanne (EPFL), 2002 Neuch\^{a}tel, Switzerland}
\newcommand{\pkh}{Planetek Hellas E.P.E., 44 Kifisias Avenue, 15125 Marousi, Athens, Greece}

\begin{document}

\title{Towards quantum 3D imaging devices: the Qu3D project}

\author{Cristoforo Abbattista}
\affiliation{\pkh}

\author{Leonardo Amoruso}
\affiliation{\pkh}

\author{Samuel Burri}
\affiliation{\epfl}

\author{Edoardo Charbon} 
\affiliation{\epfl}

\author{Francesco Di Lena} 
\affiliation{\infn}

\author{Augusto Garuccio}
\affiliation{\infn}
\affiliation{\uniba}

\author{Davide Giannella}
\affiliation{\infn}
\affiliation{\uniba}

\author{Zden\v{e}k Hradil}
\affiliation{\upo}

\author{Michele Iacobellis}
\affiliation{\pkh}

\author{Gianlorenzo Massaro}
\affiliation{\infn}
\affiliation{\uniba}

\author{Paul Mos}
\affiliation{\epfl}

\author{Libor Motka}
\affiliation{\upo}

\author{Martin Pa\'ur}
\affiliation{\upo}

\author{Francesco V. Pepe}
\affiliation{\infn}
\affiliation{\uniba}

\author{Michal Peterek}
\affiliation{\upo}

\author{Isabella Petrelli}
\affiliation{\pkh}

\author{Jaroslav \v{R}eh\'a\v{c}ek}
\affiliation{\upo}

\author{Francesca Santoro}
\affiliation{\pkh}

\author{Francesco Scattarella}
\affiliation{\infn}
\affiliation{\uniba}

\author{Arin Ulku}
\affiliation{\epfl}

\author{Sergii Vasiukov}
\affiliation{\infn}

\author{Michael Wayne}
\affiliation{\epfl}

\author{Claudio Bruschini}
\thanks{Equal last author contribution}
\affiliation{\epfl}

\author{Milena D'Angelo}
\thanks{Equal last author contribution}
\affiliation{\infn}
\affiliation{\uniba}

\author{Maria Ieronymaki}
\thanks{Equal last author contribution}
\affiliation{\pkh}

\author{Bohumil Stoklasa}
\thanks{Equal last author contribution}
\affiliation{\upo}

\begin{abstract}

We review the advancement of the research toward the design and implementation of \textit{quantum plenoptic cameras}, radically novel 3D imaging devices that exploit both momentum-position entanglement and photon-number correlations to provide the typical refocusing and ultra-fast, scanning-free, 3D imaging capability of plenoptic devices, along with dramatically enhanced performances, unattainable in standard plenoptic cameras: diffraction-limited resolution, large depth of focus, and ultra-low noise. To further increase the volumetric resolution beyond the Rayleigh diffraction limit, and achieve the quantum limit, we are also developing dedicated protocols based on quantum Fisher information. However, for the quantum advantages of the proposed devices to be effective and appealing to end-users, two main challenges need to be tackled. First, due to the large number of frames required for correlation measurements to provide an acceptable SNR, quantum plenoptic imaging would require, if implemented with commercially available high-resolution cameras, acquisition times ranging from tens of seconds to a few minutes. Second, the elaboration of this large amount of data, in order to retrieve 3D images or refocusing 2D images, requires high-performance and time-consuming computation. To address these challenges, we are developing high-resolution SPAD arrays and high-performance low-level programming of ultra-fast electronics, combined with compressive sensing and quantum tomography algorithms, with the aim to reduce both the acquisition and the elaboration time by two orders of magnitude. Routes toward exploitation of the QPI devices will also be discussed.
\end{abstract}

\maketitle

\section{Introduction}

Fast, high-resolution, and low-noise 3D imaging is highly required in the most diverse fields, from space imaging to biomedical microscopy, security, industrial inspection, cultural heritage \cite{3,4,5,6,7}. In this context, conventional plenoptic imaging represents one of the most promising techniques in the field of 3D imaging, due to its superb temporal resolution: 3D imaging is realized in a single shot, for 7 frames per second at 30M pixel resolution, and 180 frames per second for 1M pixel resolution \cite{7}; no multiple sensors, near-field techniques, time-consuming scanning or interferometric techniques are required. However, conventional plenoptic imaging entails a loss of resolution which is often unacceptable. Our strategy to break such limitation consists in combining a radically new and foundational approach with last-generation hardware and software solutions. The fundamental idea is to exploit the information stored in correlations of light by using novel sensors and measurement protocols to implement a very ambitious task: high speed (10-100 fps) quantum plenoptic imaging (QPI) characterized by ultra-low noise and an unprecedented combination of resolution and depth-of-field. The developed imaging technique aims at becoming the first practically usable and properly ``quantum'' imaging technique that surpasses the intrinsic limits of classical imaging modalities. In addition to the foundational interest, the quantum character of the technique allows extracting information on 3D images from correlations of light at very low photon fluxes, thus reducing the scene exposure to illumination.

In view of the deployment of quantum plenoptic cameras suitable for real-world applications, the crucial challenge is represented by the reduction of both the acquisition and data elaboration times. In fact, a typical complication arises in quantum imaging modalities based on correlation measurements: the reconstruction of the correlation function encoding the desired image requires collecting a large number of frames (30.000-50.000 in the first experimental demonstration of the refocusing capability of correlation plenoptic imaging \cite{2}), which must be individually read and stored before elaborating the output. Therefore, to get an estimate of the total time required to form a quantum plenoptic image, the data reading and transmission times must be added to the acquisition time of the employed sensor. This problem is addressed by an interdisciplinary approach, involving the development of ultrafast single-photon sensor systems, based on SPAD arrays \cite{Zappa,Edoardo,15,16,17,18}, the optimization of circuit electronics to collect and manage the high number of frames (e.g., by GPU) \cite{9,10}, the development of dedicated algorithms (compressive sensing, machine learning, quantum tomography) to achieve the desired SNR with a minimal number of acquisitions \cite{11,12,13,14}.

Finally, the performances of QPI will be further enhanced by a novel approach to imaging based on quantum Fisher information \cite{22,26}. Treating the physical model of plenoptic imaging in the view of quantum information theory brings new possibilities of improving the setup towards super-resolution capability in the object 3D space. Having the optimal set of optical setup parameters enables object reconstruction close to ultimate limits set by nature.

The work presented in this paper is carried within the project ``Quantum 3D Imaging at high speed and high resolution'' (Qu3D), founded by the 2019 QuantERA call \cite{QuantERA}, which involves experts from three scientific research institutions, Istituto Nazionale di Fisica Nucleare (INFN, Italy), Palacky University Olomouc / Department of Optics (UPOL, Czechia), and \'Ecole polytechnique f\'ed\'erale de Lausanne / Advanced Quantum Architecture Lab (EPFL, Switzerland), and from the industrial partner Planetek Hellas E.P.E.\ (PKH, Greece).

The paper is organized as follows: in Section 2, we discuss the working principle and recent advances of Correlation Plenoptic Imaging (CPI), a technique that represents the direct forerunner of QPI; in Section 3, we present the hardware innovations currently investigated to reduce the acquisition times in CPI; in Section 4, we review the algorithmic solutions to improve QPI; in Section 5, we outline the perspectives of our future work in the context of Qu3D project; in Section 6, we discuss the relevance of our research.

\section{Plenoptic imaging with correlations: from working principle to recent advances}

Quantum plenoptic cameras promise to offer the advantages of plenoptic imaging, primarily ultrafast and scanning-free 3D imaging and refocusing capability, with performances that are beyond reach for the classical counterpart. State-of-the-art plenoptic imaging devices are able to acquire multi-perspective images in a single shot \cite{7}. Their working principle is based on the simultaneous measurement of both the spatial distribution and the propagation direction of light in a given scene. The acquired directional information translates into refocusing capability, augmentable depth of field (DOF), and parallel acquisition of multi-perspective 2D images, as required for fast 3D imaging.

\begin{figure}
    \centering
    \includegraphics[width=\linewidth]{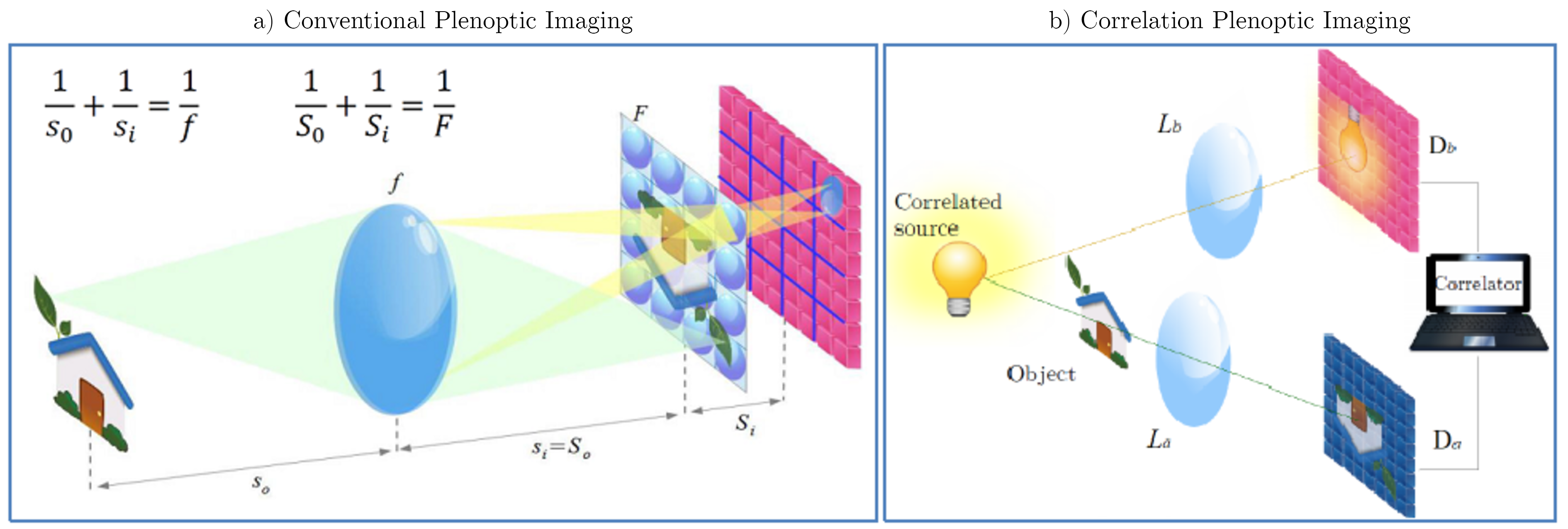}
    \caption{In panel (a), the scheme of a conventional plenoptic imaging (PI) device is shown: the image of the object is focused on a microlens array, while each microlens focuses an image of the main lens on the pixels behind. Such a configuration entails a loss of spatial resolution proportional to the gain in directional resolution.
    Panel (b) shows the scheme of a correlation plenoptic imaging (CPI) setup, in which directional information is obtained by correlating the signals retrieved by a sensor on which the object is focused with a sensor that collects the image of the light source. The image in panel (a) is reproduced with the permission from Ref.~\cite{2}, copyright American Physical Society, 2017.}
    \label{fig:plenoptic}
\end{figure}

In state-of-the-art plenoptic cameras \cite{raytrix}, directional detection is achieved by inserting a microlens array between the main lens and the sensor of a standard digital camera (see Fig.~\ref{fig:plenoptic}a). The sensor acquires composite information that allows identification of both the object point and the lens point where the detected light is coming from. However, the image resolution decreases with inverse proportionality to the gained directional information, for both structural (use of a microlens array) and fundamental (Gaussian limit) reasons; plenoptic imaging at the diffraction limit is thus considered to be unattainable in devices based on simple intensity measurement \cite{7}.

\begin{figure}
    \centering
    \includegraphics[width=\linewidth]{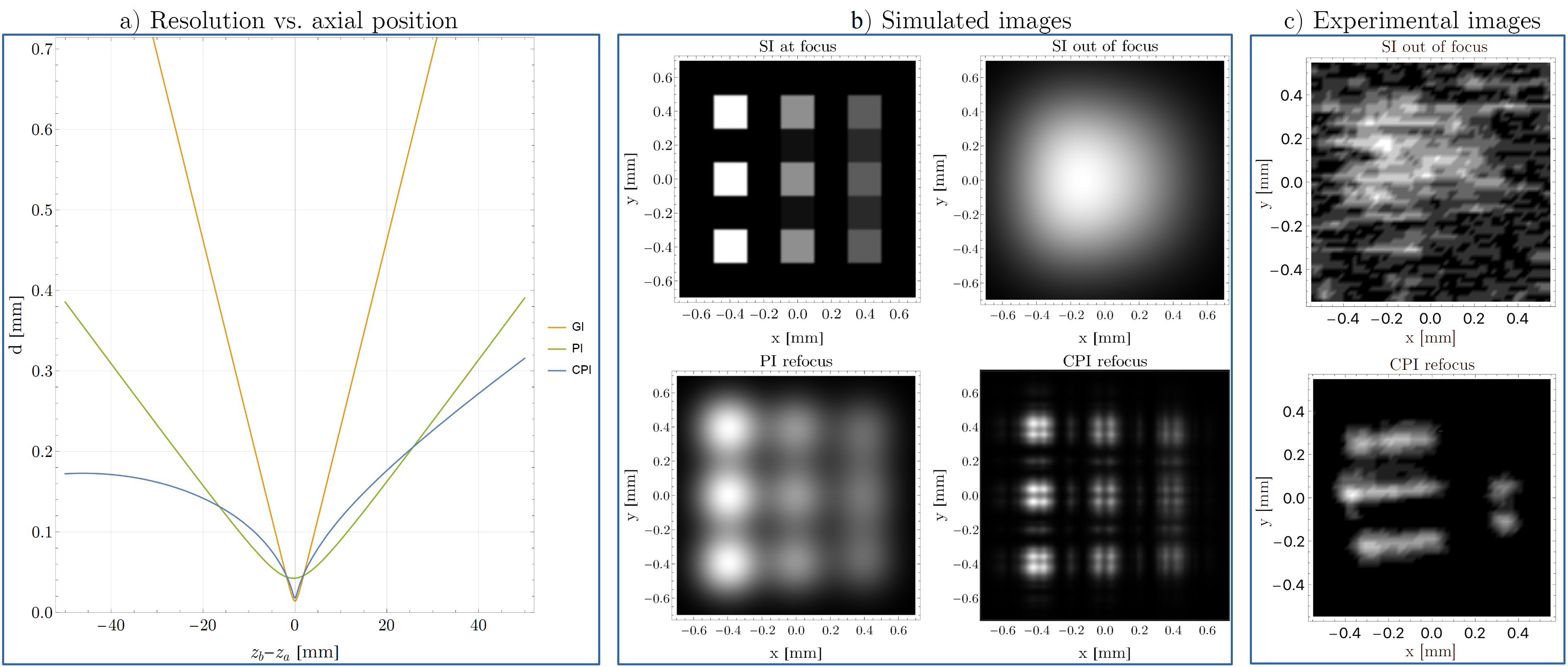}
    \caption{Panel (a) shows the resolution limits, as a function of the longitudinal position, of the image of a double-slit mask with center-to-center distance $d$ equal to twice the slit width; here, CPI outperforms both conventional imaging and standard PI with 3x3 directional resolution. 
    Plots in panel (b) show a result of a simulation: the target is moved from the focused plane (top left) to an out-of-focus plane (top right). Starting from this position we show the results of PI refocusing whit 3x3 directional resolution (bottom left) and the CPI refocusing (bottom right).
    Panel (c) shows the results of an experiment \cite{2} in which the standard image of a triple slit was completely blurred (top), while the image obtained by CPI (bottom) was made fully visible by exploiting information on light direction. Plots in panel (c) are reproduced with the permission from Ref.~\cite{2}, copyright American Physical Society, 2017.}
    \label{fig:simexp}
\end{figure}

Recently, the INFN group involved in Qu3D has proposed a novel technology, named Correlation Plenoptic Imaging (CPI), that enables to overcome the resolution drawback of current plenoptic devices, while keeping their advantages in terms of refocusing capability and 3D reconstruction \cite{1,2,3}. CPI is based on either intensity correlation measurement or photon coincidence detection, according to the light source: actually, CPI can be based on the spatio-temporal correlations characterizing both chaotic sources \cite{1,2} and entangled photon beams \cite{27} to encode the spatial and directional information on two disjoint sensors, as shown in Fig.~\ref{fig:plenoptic}b. CPI with chaotic light is based on the measurement of the correlation function
\begin{equation}
\Gamma(\bm{\rho}_a,\bm{\rho}_b) = \langle I_a(\bm{\rho}_a) I_b (\bm{\rho}_b) \rangle - \langle I_a(\bm{\rho}_a) \rangle \langle I_b (\bm{\rho}_b) \rangle,
\end{equation}
where $\langle\dots\rangle$ denotes the average on the source statistics, and $I_j\bm{\rho}_j$ ($j=a,b$) are the intensities propagated by the beam $j$ and registered in correspondence of point $\bm{\rho}_j$ on the sensor $\mathrm{D}_j$. Experimentally, the statistical averages are replaced by time averages, obtained by retrieving a collection of frames, simultaneously acquired by the two detectors. In CPI devices, the correlation function encodes combined information on the distribution of light on two reference planes, one of which corresponds to the ``object plane'' that would be focused on the sensor in a standard imaging setup, placed at a distance $s_o$ from the focusing element. In general, given an object placed at a distance $s$ from the focusing element, and characterized by the light intensity distribution $\mathcal{A}(\bm{\rho})$, its images are encoded in the function $\Gamma(\bm{\rho}_a,\bm{\rho}_b)$, in the geometrical-optics limit, as
\begin{equation}
\Gamma(\bm{\rho}_a,\bm{\rho}_b) \sim \mathcal{A}^n \left( \frac{s}{s_o} \frac{\bm{\rho}_a}{M} + \left( 1 - \frac{s}{s_o} \right) \frac{\bm{\rho}_b}{M_L} \right) ,
\end{equation}
where $M$ and $M_L$ are the magnifications of the images of the reference object plane and of the focusing element, respectively, while the power $n$ is equal to 1 or 2, according to whether the object lies in only one \cite{2,1,28} or both \cite{scagliola,cpiap} optical paths.

Experimental CPI based on pseudo-thermal light is shown in Fig.~\ref{fig:simexp}c, where both the acquired out-of-focus image and the corresponding refocused image are shown \cite{2}. It was demonstrated in this proof of principle that CPI is characterized by diffraction-limited resolution on the object plane focused on the sensor. Details on the resolution limits are shown in Fig.~\ref{fig:simexp}a, where one can observe an even more striking effect: thanks to its intrinsic coherent nature, CPI enables an unprecedented combination of resolution and DOF \cite{2}. However, the low-noise sCMOS camera employed in the experiment, working at 50 fps at full resolution, requires several minutes to acquire 30,000 frames used to reconstruct the plenoptic correlation function, and a standard workstation has taken over 10 hours for elaborating the acquired data and perform refocusing. The resulting image was also rather noisy, due to the well-known resolution vs. noise compromise of chaotic light ghost imaging, that keeps affecting also CPI \cite{28}.


We are addressing these issues by employing two kinds of sources:
\begin{itemize}
\item Chaotic light sources, such as pseudothermal light, natural light, LEDs and gas lamps, and even fluorescent samples, operated either in the high-intensity regime or in the ``two-photon'' regime, in which an average of 2 photons per coherence area propagates in the setup. Chaotic light sources are well-known to be characterized by EPR-like correlations in both momentum and position variables \cite{laserphyslett,gatti}, to be exploited in an optimal way to retrieve an accurate plenoptic correlation function in the shortest possible time. In order to efficiently retrieve spatio-temporal correlations, tight filtering of the source can be necessary to match the filtered source coherence time with the response time of the SPAD arrays, that can be as low as 1 ns. Alternatively, pseudorandom sources with a controllable coherence time, made by impinging laser light on a fast-switching digital micromirror device (DMD), can be employed. Interestingly, recent studies have shown that, in the case of chaotic light illumination, the plenoptic properties of the correlation function do not need to rely strictly on ghost imaging: correlations can be measured between any two planes where ordinary images (see Fig.~\ref{fig:plenoptic}b) are formed \cite{28}. This discovery has led to the intriguing result that the SNR of CPI improves when ghost imaging of the object is replaced by standard imaging \cite{29}. In particular, excellent noise performances are expected in the case of images of birefringent objects placed between crossed polarizers. This kind of source is particularly relevant in view of applications in fields like biomedical imaging (cornea, zebrafish, invertebrates, biological phantoms such as starch dispersions), security (distance detection, DOF extension), and satellite imaging.
\item Momentum-position entangled beams, generated by spontaneous parametric down-conversion (SPDC), which have the potential to combine QPI with sub-shot noise imaging \cite{30}, thus enabling high-SNR imaging of low-absorbing samples, a challenging issue in both biomedical imaging and security.
\end{itemize}
The design of both quantum plenoptic devices are currently undergoing optimization by implementing a novel protocol that enables to further mitigate the resolution vs DOF compromise with respect to the one shown in Fig.~\ref{fig:simexp}a: this protocol is based on the observation that, for any given resolution, the DOF can be maximized by correlating the standard images of two arbitrary planes, chosen in the surrounding of the object of interest, instead of imaging the focusing element \cite{cpiap}. Moreover, we are investigating the possibility to merge quantum plenoptic imaging with the measurement protocols developed in the context of differential ghost imaging \cite{differential}.

\section{Hardware speedup: advanced sensors and ultra-fast computing platforms}

To improve the performances of CPI in terms of acquisition speed and data elaboration time, we are employing dedicated advanced sensors and  ultra-fast computing platforms. In this section, we describe the details of the implementation and the perspectives on these fields.

\begin{figure}
    \centering
    \includegraphics[width=0.6\linewidth]{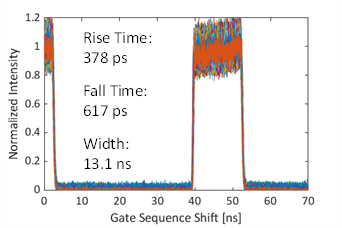}
    \caption{SwissSPAD2 gate window profile. The transition times and the gate width are annotated in the figure. The gate width is user-programmable, and the minimum gate width in the internal laser trigger mode is $10.8$ ns.}
    \label{fig:gate_window}
\end{figure}

\subsection{SPAD arrays as high-resolution time-resolved sensors}

A relevant part of the speedup that we are seeking is determined by replacing commercial high-resolution sensors, like scientific CMOS and EMCCD cameras, with sensors based on cutting-edge technology such as single-photon avalanche diode (SPAD) arrays.
A SPAD is basically a photodiode which is reversely biased above its breakdown voltage, so that a single photon which impinges onto its photosensitive area can create an electron-hole pair, triggering in turn an avalanche of secondary carriers and developing a large current on a very short timescale (picoseconds) \cite{Zappa,Edoardo}. This operation regime is known as Geiger mode. The SPAD output voltage is sensed by an electronic circuit and directly converted into a digital signal, further processed to store the binary information that a photon arrived, and/or the photon time of arrival. In essence, a SPAD can be seen as a photon-to-digital conversion device with exquisite temporal precision. SPADs can also be gated, in order to be sensitive only within temporal windows as short as a few nanoseconds, as shown in Fig.~\ref{fig:gate_window}. Individual SPADs can nowadays be used as the building blocks of large arrays, with each pixel circuit containing both the SPAD and the immediate photon processing logic and interconnect. Several CMOS processes are readily available and allow to tailor both the key SPAD performance metrics and the overall sensor or imager architecture \cite{LSA,Caccia}. Sensitivity and fill-factor have for some time lagged behind those of their scientific CMOS or EMCCD counterparts, but have been substantially catching up in recent years.

\begin{figure}
    \centering
    \includegraphics[width=0.25\linewidth]{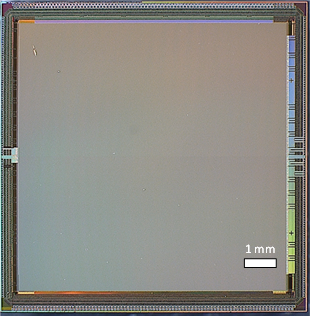}
    \includegraphics[width=0.65\linewidth]{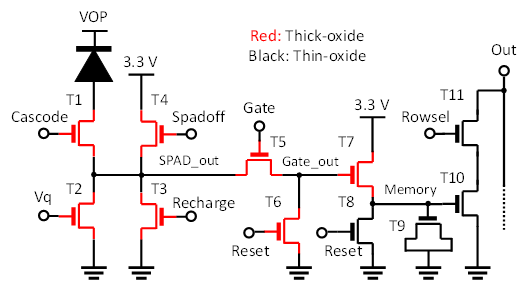}
    \caption{SwissSPAD2 photomicrograph (left) and pixel schematics (right). The pixel consists of 11 NMOS transistors, 7 with thick-oxide and 4 with thin-oxide gate. The pixel stores a binary photon count in its memory capacitor. The in-pixel gate defines the time window, with respect to a 20 MHz external trigger signal, in which the pixel is sensitive to photons.}
    \label{fig:swissspad2}
\end{figure}

Based on the requirements of QPI, we have chosen to employ the SwisSPAD2 array developed by the AQUA laboratory group of EPFL, characterized by a 512x512 pixel resolution (see Fig.~\ref{fig:swissspad2}), which is one of the widest and most advanced SPAD arrays to date \cite{16,18}. The sensor is internally organized as two halves of 256x512 pixels to reduce load and skew on signal lines and enable faster operation. It is a purely binary gated imager, i.e. each pixel records either a 0 (no photon) or a 1 (one or more photons) for each frame, with basically zero readout noise. The sensor is controlled by an FPGA generating the control signals for the gating circuitry and readout sequence and collecting the pixel detection results. In the FPGA, the resulting one-bit images can be further processed, e.g. accumulated into multi-bit images, before being sent to a computer/GPU for analysis and storage. The maximum frame rate is 97.7 kfps, and the native fill factor of 10.5\% can be improved by 4-5 times, for collimated light, by means of a microlens array \cite{17} (higher values are expected from simulation after optimization); the photon detection probability is 50\% (25\%) at 520 nm (700 nm) and 6.5 V excess bias. The device is also characterized by low noise (typically less than 100 cps average Dark Count Rate per pixel at room temperature, with a median value about 10 times lower) and advanced circuitry for nanosecond gating. A detailed comparison of SwissSPAD2 with other large-format CMOS SPAD imaging cameras is presented in Ref.~\cite{18}.

Currently, we are using the available version of SwissSPAD2, with a 512x256 pixel resolution, to generate sequences of frames and store them in an on-board 2 GB memory, before transferring them to a computer by a standard USB3 connection, which can be done using existing hardware. We are integrating this sensor in the prototype of chaotic-light base quantum plenoptic camera in a way that two disjoint halves of the sensor (of 256x256 pixels each) are used for retrieving the images of the two reference planes. The high speed of this sensor is expected to reduce the acquisition time of quantum plenoptic images by 2 order of magnitudes with respect to the first CPI experiment \cite{2}, in which the region of interest on the sensor was made of $700\times 700$ pixels for the spatial measurement side, and $600\times 600$ for the directional measurement side; $2\times 2$ binning on both sides during acquisition and a subsequent $10\times 10$ binning on the directional side led to the effective spatial resolution of $350\time 350$ pixels, and angular resolution of $25\times 25$ pixels.

We are also working towards several further optimizations of the sensor system, e.g. by developing gating for noise reduction in QPI devices. In order to employ the full sensor (512x512 pixels), we are implementing a synchronization mechanism for a pair of imagers by means of two FPGAs, so as to operate on a common time-base at the nanosecond level (to this end, two control circuits shall operate from a single clock and have a direct communication link). Finally, the SwissSPAD2 arrays are being integrated with a fast communication interface in order to speed up data transfer and make it possible to deliver, in a sustained way, full binary frame sequences to a GPU. The latter will run advanced algorithms for data pre-processing, image reconstruction and optimization, as we will discuss in more detail in the next subsection.

\subsection{Computational hardware platform}

In a QPI device integrated with SwissSPAD2, the acquired data rate for a single frame acquisition can be estimated to 26 Gb/s, which is beyond the reach of standard data buses. This poses great challenges in both hardware and software design. Our approach is to employ the expertise of PKH to seek careful design of electronic interconnections (buses) between sensor control electronics and processing device, theoretical refinement and optimization of algorithms (e.g., compressive sensing \cite{19,20}), porting to an efficient computational environment, and design of a specific acquisition electronics for optimizing data flow from the light sensors to a dedicated processing system, able to guarantee the required computational performance (e.g., exploiting GPUs or FPGA) \cite{9,10}.

The introduction of an embedded data acquisition- and processing board, integrating a GPU, aims at data pre-processing, thus significantly reducing the amount of data to be transferred to (and saved on) an external workstation. GPUs exploit a highly parallel elaboration paradigm, enabling to design algorithms that run in parallel on hundreds or thousands of cores and to make them available on embedded devices. A great advantage of GPUs is programmability: many standard tools exist (e.g. OpenCL and CUDA) that allow fast and efficient design of complex algorithms that can be injected on the fly in the GPU memory for accomplishing tasks ranging from simple filtering to advanced machine learning. Efforts are made to design a \textit{processing matrix}, so that each line and column of the sensor will be managed by a dedicated portion of the heterogeneous processing platform (CPU/GPU/FPGA): the \textit{pixel series processor}. Those dedicated units will be interconnected to one another to cooperate for implementing algorithms that require distributed processing on a very small scale.

The embedded acquisition-processing board is designed to best fit to SPAD array and SW application needs. The optimal system design will be evaluated by considering theoretical algorithms, engineered SW implementations, HW set-up and HW/SW trade-off. We will identify a preliminary set of possible configurations and perform a trade-off by comparing overall performances, considering the requirements in data quality, processing speed, costs, complexity, etc.

\begin{figure}
\centering
\includegraphics[width=0.9\linewidth]{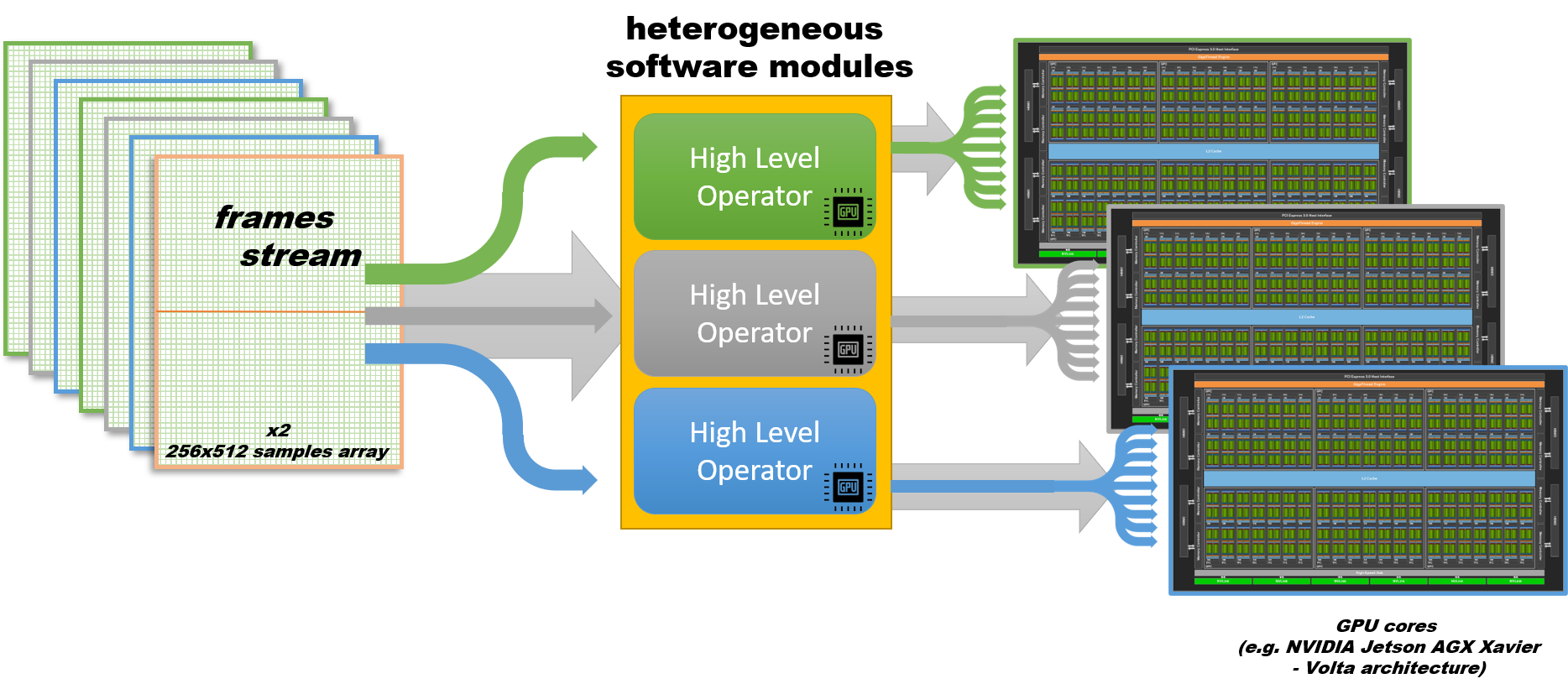}
\caption{Qu3D scenario for GPU parallel processing based on NVIDIA Volta architecture as provided by NVIDIA Jetson AGX Xavier device.}\label{fig:jetson}
\end{figure}

Based on the challenging objectives to be achieved, a preliminary analysis based on COTS (Commercial-Off-The-Shelf) solutions was performed, in order to identify a set of accelerating devices addressing QPI requirements in terms of computing capability and portability. The option offered by the NVIDIA Jetson Xavier AGX board shown in Fig.~\ref{fig:jetson} is considered a promising candidate to achieve our goal. This device indeed offers an encouraging performance/integration ratio with low power usage and very interesting computing capabilities. Its main characteristics are reported in Table ~\ref{tab:jetson}.

 \begin{table}
 \caption{NVIDIA Jetson Xavier AGX: Technical Specifications.\label{tab:jetson}}
 \begin{tabular}{ccc}
 \toprule
 \textbf{}	& \textbf{Jetson Xavier AGX}	\\ \hline
GPU			& 512-core NVIDIA Volta$^{\text{TM}}$ GPU with 64 Tensor Cores\\
CPU			& 8-core NVIDIA Carmel Arm\textsuperscript{\textregistered} v8.2 64-bit CPU 8MB L2 + 4MB L3\\
Memory		& 16 GB 256-bit LPDDR4x 136.5GB/s\\
PCIE			& 1 x8 + 1 x4 + 1 x2 + 2 x1 (PCIe Gen4, Root Port \& Endpoint)\\
DL Accelerator	& 2x NVDLA Engines\\
Vision Accelerator	& 7-Way VLIW Vision Processor\\
Connectivity		& 10/100/1000 BASE-T Ethernet\\ \hline
 \end{tabular}
 \end{table}

Considering the listed HW capabilities, despite the ARM processor has a limited computing power when compared to high-end desktop processors, it allows to leverage multi-core capabilities for implementing that part of the code that will feed the quite powerful GPU device on board. In addition, the foreseen optimization strategy will require a dedicated implementation able to consider the maximum amount of memory of 16GiB available, shared with the GPU. Also, the solution to be developed should take into account the bandwidth of about 136GB/s of the on-board memory, which may represent a limiting factor when GPU and CPU exchange buffers. An implementation based on the CUDA framework -over OpenCL or other technologies- will be preferred to best use the NVIDIA device.
Finally, given the capability of the NVDLA Engines to perform multiplications and accumulations in a very fast way, we consider it interesting to perform an assessment of how to exploit these devices for implementing the QPI-specific correlation functions and/or other multiplication/sum intensive computations.

Along with the assessment of a dedicated HW solution, further optimizations applicable at the algorithmic level have been considered in order to enrich the engineered device with a highly customized algorithmic workflow able to exploit the peculiarities of the CPI technique and its related input data. More specifically, we are analysing those steps of QPI processing that appear as more computationally demanding, thus representing a bottleneck for performances. To this end, tailored reshaping operations applied over the original 3-dimensional multi-frame structure of the input data were explored, to facilitate the development of a parallelized elaboration paradigm for evaluating the CPI-related correlation function. Besides, the peculiar feature of the input dataset to be acquired by SPAD sensors as one-bit images will be valued through dedicated implementations able to gain from intensive multiplication/sum math among binary-valued variables.

\section{Quantum and classical image-processing algorithms}

Further reduction of the acquisition speed and the optimization of the elaboration time is addressed by exploiting dedicated quantum and classical image processing, as well as novel mathematical methods coming on one hand from compressive sensing, on the other hand from quantum tomography and quantum Fisher information.

\subsection{Compressive sensing}

In order to reduce the amount of required data (at present $10^3$--$10^4$ frames) by at least 1 order of magnitude, we are exploring different approaches. First, we are investigating the opportunity of implementing compression techniques for improving bus bandwidth utilization, thus acting on data transfer optimization. Data compression may also rely on manipulation of raw input data to determine only the relevant information in a sort of information bottleneck paradigm, in which software nodes in a lattice provide their contribution to a probable reconstruction of the actual decompressed data. This is very similar to artificial neural network structures, where the network stores a representation of a phenomenon and returns a response based on some similarity rating versus the observed data. Moreover, the availability of advanced processing technologies allows the investigation and implementation of alternative techniques able to exploit the sparsity of the retrieved signal to reconstruct information from a heavily sub-sampled signal, by compressive sensing techniques \cite{19,20}.

As in other correlation imaging techniques, such as ghost imaging (GI), in the CPI protocol the object image is reconstructed by performing correlation measurements between intensities at two disjoint detectors. Katz et al. \cite{Katz09} demonstrated that conventional GI offers the possibility to perform compressive sensing (CS) boosting the recovered image quality. 

CS theory asserts the possibility of recovering the object image from far fewer samples than required by the Nyquist–Shannon sampling theorem and it relies on two main principles: sparsity of the signal (once expressed in a proper basis) and incoherence between the sensing matrix and the sparsity basis. 

In conventional GI, the transmission measured for each speckle pattern represents a projection of the object image and CS finds the sparsest image among all the possible images consistent with the projections. In practice, CS reconstruction algorithms solve a convex optimization problem, seeking for the image which minimizes the $L_1-$norm in the sparse basis among the ones compatible with the bucket measurements, see Refs.\cite{Jiying10,Abmann13,Chen19, Liu20} for a review. 

\begin{figure}
\includegraphics[width=\linewidth]{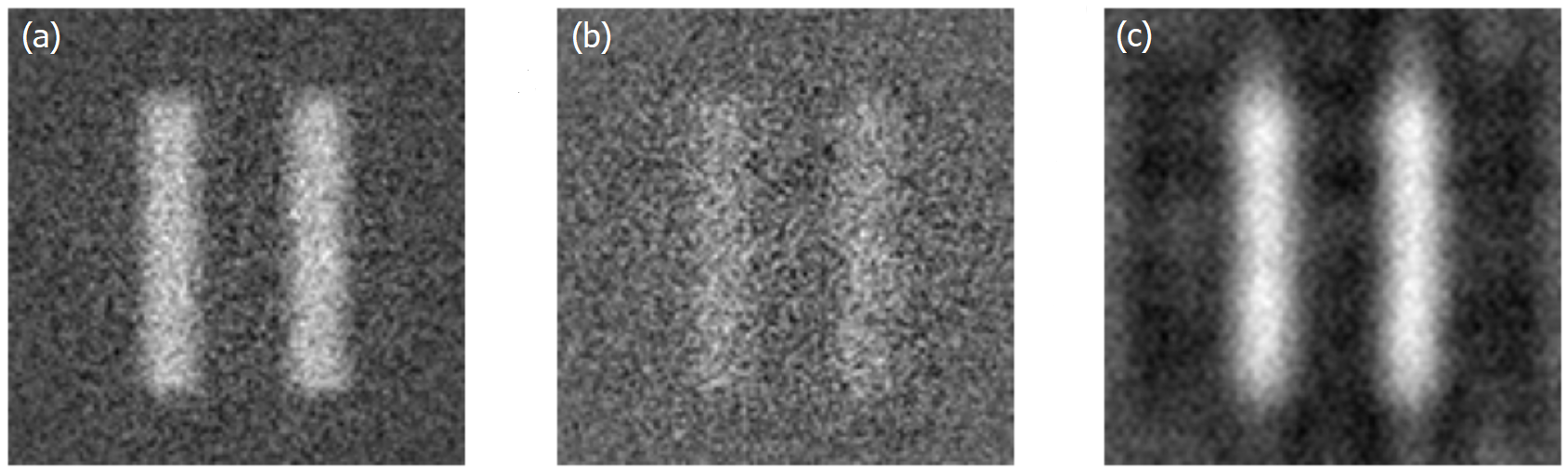}
\centering
\caption{Panel (a): double-slit image reconstruction obtained by correlations measurements, considering $N=6000$ frames and two detectors characterized by a $128\times 128$ and a $10\times 10$ pixel resolution. Panel (b): the standard reconstruction is repeated considering only the $10\%$ of the available frames, chosen randomly. Panel (c): compressive sensing reconstruction using the same dataset as in panel (b). While in the first case the Pearson's correlation coefficient is $r_{red}=0.55$, in the latter case the coefficient is increased to $r_{CS}=0.81$.}
\label{fig:CS}
\end{figure}

We are developing a novel protocol reducing the number of measurements required for image recovery by an order of magnitude. Once properly refocused, a single acquisition can be fed into the compressive sensing algorithm several times thus exploiting the plenoptic properties of the acquired data and increasing the signal-to-noise ratio of the final refocused image. We tested the CS-CPI algorithm with numerical simulations, as summarized in Fig.~\ref{fig:CS}. In panel (a) a double-slit mask is reconstructed by correlations measurements considering $N=6000$ frames, in panel (b) the standard reconstruction is repeated considering only the $10\%$ of available frames, while in panel (c) the CS reconstruction using the same reduced set of measured data. In addition to a data-fidelity term corresponding to a linear regression, we penalized the $L_1-$norm of the reconstructed image to account for its sparsity in the $2D-DCT$ domain. The resulting optimization problem is known as the LASSO (Least Absolute Shrinkage and Selection Operator) \cite{lasso}. We employed the coordinate descent algorithm to efficiently solve it and we set the regularization parameter, controlling the degree of sparsity of the estimated coefficients, by cross-validation. In this proof-of-concept experiment, we simply use Pearson's correlation coefficient to measure the similarity between the reconstructions obtained using the restricted dataset and the image obtained considering all the $N=6000$ frames.

\subsection{Plenoptic tomography}
Novel reconstruction algorithm with the advantage of real 3D image lack of artifacts is based on the idea of recasting plenoptic imaging as an absorption tomography. The classical refocusing algorithm is based on superimposing images of the 3D scene from different viewpoints, which necessarily results into the contribution of out-of-focus parts of the scene, the effect responsible for the existence of blurred part of 3D image reconstruction. The tomography approach is based on a different principle and provides sufficient axial and transversal resolution without artifacts.
For this purpose, the object space is divided into voxels and the goal is to reconstruct absorption coefficients of each voxel. The measured correlation coefficient of a pair of points from the correlated planes is transformed to be linearly proportional to the attenuation along the ray path connecting those two points. Classical inverse Radon transform can be used in this scenario to obtain a 3D image but using the Maximum Likelihood absorption tomography algorithm \cite{MaxLikTomo} further enhances the quality of the tomography reconstruction and performs well even for a small range of projections angels. In fact, advanced tomographic reconstruction algorithms based on the Maximum Likelihood principle are more resistant to noise and require fewer acquisitions, for a given precision, in comparison to standard tomographic protocols. We are investigating special tools to deal with informationally incomplete detection schemes for very high resolution, and optimal methods for data analysis based on convex programming tools.

\subsection{Quantum tomography and quantum Fisher information}

A further quantum approach to image analysis and detection schemes will be employed to achieve super-resolution (or, eventually, for maintaining the desired resolution and speeding up acquisition and elaboration of the quantum plenoptic images) and to compare correlation plenoptic detection scheme to the ultimate quantum limits. The basic concept underpinning the Fisher Information super-resolution imaging is the formal mathematical analogy between the classical wave optics and quantum theory, which makes it possible to apply the advanced tools of quantum detection and estimation theory to classical imaging and metrology \cite{21,22}. The dvantage of this approach lies in the ability to quantify the performance of the imaging setup based on rigorous statistical quantities. Inspired by the quantum theory of detection and estimation, quantum Fisher information, quantity connected with the ultimate limits allowed by nature, is computed for simple 3D imaging scenarios like localization and resolution of two points in the object 3D space  \cite{23,24,25,26}. For example, one might be interested in measuring the separation of two point-like sources and seek the optimal detection scheme extracting the maximum amount of information about this parameter. We shall thus employ quantum Fisher information to design optimal measurement protocols within the quantum plenoptic devices, able to extract specific relevant information, for enhanced resolution, with a minimal number of acquisitions. The question of the minimal number of detections over the correlation planes which achieves acceptable reconstruction quality is of relevance because of direct connection to the amount of processed data and can lead to reduced time for data processing.

\section{Perspectives}

In the context of the Qu3D project, all the developments and technologies presented in the previous sections will be integrated into the implementation of two quantum plenoptic imaging (QPI) devices, namely,
\begin{itemize}
    \item a compact single-lens plenoptic camera for 3D imaging, based on the photon number correlations of a dim chaotic light source;
    \item an ultra-low noise plenoptic device, based on the correlation properties of entangled photon pairs emitted by spontaneous parametric down-conversion (SPDC), enabling 3D imaging of low-absorbing samples, at the shot-noise limit or below.
\end{itemize}
Our objective is to achieve in both devices high resolution, whether diffraction-limited or sub-Rayleigh, combined with a DOF larger by even one order of magnitude compared to standard imaging. The science and technology developed in the project will contribute to establishing a solid baseline of knowledge and skills for the development of a new generation of imaging devices, from quantum digital cameras enhanced by refocusing capability to quantum 3D microscopes \cite{scagliola} and space imaging devices.

\section{Conclusions}

We have presented the challenging research directions we are following to achieve practical quantum 3D imaging: minimizing the acquisition speed without renouncing to high SNR, high resolution and large DOF \cite{3,4,5,6,7}. Our work represents a significant advance with respect to the state-of-the-art of both classical and quantum imaging, as it enhances the performances of plenoptic imaging and dramatically speeds up quantum imaging, thus facilitating the real-world deployment of quantum plenoptic cameras.

This ambitious goal will be facilitated by working toward the extension of the reach of quantum imaging to other fields of science, and opening the way to new opportunities and applications, including the prospects of offering new medical diagnostic tools, such as 3D microscopes for biomedical imaging, as well as novel devices (quantum digital cameras enhanced by 3D imaging, refocusing and distance detection capabilities) for security, space imaging, and industrial inspection. The collaboration crosses the traditional boundaries between the involved disciplines: quantum imaging, ultra-fast cameras, low-level programming of GPU, compressive sensing, quantum information theory, and signal processing.

\section*{Acknowledgments}
Project Qu3D is supported by the Italian Istituto Nazionale di Fisica Nucleare, the Swiss National Science Foundation (grant 20QT21 187716 ``Quantum 3D Imaging at high speed and high resolution''), the Greek General Secretariat for Research and Technology, the Czech Ministry of Education, Youth and Sports, under the QuantERA programme, which has received funding from the European Union's Horizon 2020 research and innovation programme.

\end{document}